# Enhancement of $J_c$ by Hf -Doping in the Superconductor $MgB_2$: A Hyperfine Interaction Study

**A.Talapatra[1], S.K.Das[1], S.K.Bandyopadhyay[1*], P.Barat[1], Pintu Sen[1], R.Rawat[2], A.Banerjee[2], T. Butz[3]**

1.Variable Energy Cyclotron Centre, 1/AF, Bidhan Nagar, Kolkata-700 064, India.

2.UGC-DAE Consortium for Scientific Research, Khandwa Road, Indore, 452 017.

3. Universität Leipzig, Fakultät für Physik und Geowissenschaften, Institut für Experimentelle Physik II, Linnéstraße 5, 04103 Leipzig, Germany

**Abstract:**

Measurements of the critical current density ($J_c$) by magnetization and the upper critical field ($H_{c2}$) by magnetoresistance have been performed for hafnium-doped $MgB_2$. There has been a remarkable enhancement of $J_c$ as compared to that by ion irradiation without any appreciable decrease in $T_c$, which is beneficial from the point of view of applications. The irreversibility line extracted from $J_c$ shows an upward shift. In addition, there has been an increase in the upper critical field which indicates that Hf partially substitutes for Mg. Hyperfine interaction parameters obtained from time differential perturbed angular correlation (TDPAC) measurements revealed the formation of HfB and $HfB_2$ phases along with the substitution of Hf. A possible explanation is given for the role of these species in the enhancement of $J_c$ in $MgB_2$ superconductor.

***Corresponding author; e-mail: skband@veccal.ernet.in**

## Introduction:

Disorder in type II superconductors plays important roles like pinning the magnetic flux lines in the mixed phase. Pinning of these flux lines increases the critical current density ($J_c$) and irreversibility field [$H_{irr}(T)$]. With the introduction of disorder, there are changes in both microscopic and macroscopic properties. Disorder and defects in the superconducting materials alter the thermodynamic properties like the upper critical field ($H_{c2}$) and the coherence length ($\xi$). Different methods are used to enhance $J_c$ and thermodynamic properties like defect creation by ion irradiation, grain refinement, element doping etc. We had earlier studied $J_c$ and $H_{c2}$ of bulk polycrystalline $MgB_2$ irradiated with Ne ions. There was an enhancement of $H_{c2}$, but the change in $J_c$ was marginal [1]. An alternate way to introduce defects is by introducing impurity atoms of different valence states than the host ions. There are reports of Ti and Zr doping in $MgB_2$ [2]. These group IV elements on $Mg^{+2}$ substitutional site for example introduce additional charge (+2) in the cation site. It is highly probable that dopants go either to Mg site or to B site and cause an enhancement of $H_{c2}$. However, it is not clearly known whether the dopants are substituted at Mg or B positions. Hence it would be very important to have an idea of the site occupancies of the dopant to specify the roles of different species in the enhancement of $J_c$, $H_{irr}$ and $H_{c2}$. Hyperfine interaction techniques like time differential perturbed angular correlation (TDPAC) offer a great potential to understand the nature of sites occupied by the dopant which should be a nuclear probe for this technique. This technique has atomic scale resolution and is extremely sensitive to trace quantities of probe atoms. TDPAC essentially measures the hyperfine parameters, i.e. the electric field gradient (EFG) and asymmetry parameter $\eta$. Both these parameters are sensitive to the electron distribution around the probe nucleus. The measurement of the EFG thus can help telling us the kind of the site the probe is occupying. In other words, it identifies the species formed by the probe atom. In this paper, we have analysed the change in the upper critical field ($H_{c2}$), $J_c$ and $H_{irr}$ of Hf-doped $MgB_2$ as compared to the undoped one. We provide here the plausible reasons for the enhancement of $H_{c2}$ and $J_c$ by interpreting the parameters from TDPAC and X-ray diffraction (XRD) measurements. $^{181}$Hf, which can be obtained by neutron irradiation of the Hf-doped $MgB_2$, is the nuclear probe for the TDPAC measurement in this work. We have employed only 1% Hf, as it is difficult to introduce dopant in higher concentration without causing a severe reduction in $T_c$, since $MgB_2$ with its unique structure is extremely sensitive to solid solubility of other ions with different atomic size and valency [3]. As a reference material, we have employed Hf-boride synthesised under identical conditions to that of Hf-doped $MgB_2$. This was done to identify any $HfB_x$ species formed in the Hf-doped $MgB_2$.

**Experimental:**

Hf-doped $MgB_2$ and $HfB_x$ bulk samples were prepared using solid-state reaction. Mg and B powders were taken stoichiometrically with 1 atomic percent of Hf powder. They were mixed thoroughly and pelletized. The pellets were wrapped in a Ta foil and put in a quartz tube which was then evacuated and sealed in He atmosphere and put in a furnace preheated at 1073K and heated for two hours and then at 1173K for another two hours [4]. The tube was then furnace cooled. The samples were characterized by XRD taken on a Philips PW 1710 diffractometer with Cu $K_\alpha$ of wavelength 1.54Å. The resistivity and $T_c$ of the samples were measured using the four-probe technique [4]. The same procedure was followed for the case of $HfB_x$ except that the temperatures employed were 1173K and 1373K to check the extent of formation of other compounds of Hf and B like HfB or $HfB_2$.

$H_{c2}$ was obtained from magnetoresistance measurements of samples in a Philips 8T cryostat. The magnetisation measurements were carried out at 14 T QD PPMS VSM at different temperatures.

Hf-doped $MgB_2$ and $HfB_x$ were neutron activated to produce $^{181}Hf$ which decays to $^{181}Ta$ in the samples. These samples were annealed at 1173K after irradiation for 4 hours to remove any defect introduced by the neutron activation and counted for the coincidence of the 131-482 keV cascade of $^{181}Hf$ using a TDPAC camera [5] consisting of six $BaF_2$ detectors coupled to a fast-slow coincidence set up.

**Results and Discussions:**

Figs. 1(a, b) show the XRD patterns of Hf doped $MgB_2$ annealed at 1173K and $HfB_x$ annealed at 1173K. The XRD pattern in Fig. 1a clearly reveals the similarity to pure $MgB_2$. It might be due to the fact that at the level of doping employed (1%), the formation of phases containing Hf-boride or substitution of Hf in B or Mg sites is not detectable by XRD. Fig. 1b reflects the formation of multiphase $HfB_x$ with the peaks being indexed with the reflection planes corresponding to both HfB and $HfB_2$. Thus, it may be surmised that, at the temperature of formation (1173K) for Hf-doped $MgB_2$ (temperature needed for the formation of $MgB_2$), there is a mixture of HfB and $HfB_2$.

The transition temperatures $T_c^{(R=0)}$ of undoped and doped samples are 38.0K and 36.4K respectively, with transition widths ($\Delta T_c$) of 0.5K and 1.9K, respectively, as displayed in Fig. 2. Thus, $T_c^{(onset)}$ for the undoped sample is 38.5K and that of the doped sample is 38.3K. There is little change in $T_c^{(onset)}$ due to 1% Hf doping. However, here is an increase in resistivity and $\Delta T_c$ of the doped sample. The residual resistivity of the undoped sample as obtained by extrapolation is 19.0 μΩ-cm and that of the doped sample is 78.8 μΩ-cm. The doping in $MgB_2$ generally causes a decrease in $T_c$ and increase in resistivity [6-8] excepting for the MgO addition where there is an enhancement of Tc [9].

$MgB_2$ is a multiband superconductor. It has four bands at the Fermi surface ($E_F$). Two of them, σ-bands (bonding and antibonding) are formed mainly by the $p_x$+$p_y$ electrons of boron and the other two π-bands (bonding and antibonding) formed mainly by $p_z$ electrons of boron. The origin of superconductivity lies in the σ-bands. These bands are not saturated and the hole on the top of σ-band couples with $E_{2g}$ phonons to give rise to superconductivity. These bands are two dimensional (2D) in character while the π-bands are three dimensional (3D). $T_c$ of $MgB_2$ is closely proportional to the density of states of σ-bands [10]. Kortus et al. found that the $T_c$ drop is associated to band filling and interband scattering [11].

Fig. 3 shows the upper critical fields of the samples. There is an increase in $H_{c2}$ due to Hf-doping. We have fitted the $H_{c2}(T)$ curve with phenomenological equation:

$$H_{c2}(T) = H_{c2}(0)\left\{1 - \left(\frac{T}{T_C}\right)^2\right\}^{\alpha} \qquad (1)$$

This gives $H_{c2}(0)$ as 18.7 T and 24.2 T for undoped and Hf doped $MgB_2$ respectively. The enhancement is quite large as compared to that by Neon irradiation [1]. The increase in $H_{c2}$ in the doped sample is due to Hf doping. Enhancement of $H_{c2}$ was observed by Braccini et al. [12] by carbon doping and He ion irradiation on $MgB_2$ and they explained the enhancement by disorder introduced through doping. In multiband $MgB_2$, there is no correlation between the enhancement of $H_{c2}$ and the resistivity change due to doping or ion irradiation [13]. Superconducting parameters like $H_{c2}$ and $T_c$ are governed by coupling of carriers in the 2D σ-band [14], whereas conductivity is primarily controlled by the scattering in the 3D π-band [13]. The appreciable increase in $H_{c2}$ is indicative of some substitution of Hf in Mg or B sites. But, substitution in B site is likely to cause a damage to the coupling of carriers in the 2D σ-band and hence a drastic decrease in $T_c^{(onset)}$ which is absent here. Hf, if at all, is likely to be substituted dominantly in Mg site which affects the scattering in the 3D π-band and, hence the conductivity. The increase in resistivity of the doped sample is an indication of Hf substitution in Mg site as well as a decrease in grain connectivity [15]. The temperature dependence of $H_{c2}(T)$ is described well by a theory of dirty two-gap superconductivity.

The critical current density $J_c$ was obtained using Bean's critical state model:

$$J_c = \frac{20\,\Delta M}{V\,a\left(1-\frac{a}{3b}\right)} \qquad (2)$$

$\Delta M = (M_+ - M_-)$, where $M_\pm$ is the magnetic moment obtained from isothermal magnetisation measurements. The measurements were carried out at different fields. *a* and *b* are widths of the samples. *V* is the volume of the samples. The measurements were also carried out at different temperatures. Figs. 4a and 4b show the variations of $J_c$ of both pristine and Hf-doped $MgB_2$ at 2K and 20K. In both cases, the enhancements of $J_c$ for Hf doped $MgB_2$ are quite large as compared to that by Ne irradiation [1]. At the same time, the decrease of $J_c$ with applied magnetic field is lower than the undoped sample.

The irreversibility fields ($H_{irr}$) for the samples were obtained using the criteria that it is that field where the critical current density, $J_c = 100$ A/cm$^2$. The irreversibility lines (IL) of the samples are shown in the inset of Fig. 3. There is an upward shift in the IL in case of the doped sample. The IL was found to scale with temperature in the same way as $H_{c2}$. $H_{irr}(0)$ for the undoped and doped samples were obtained as 9T and 13.4T, respectively. Thus doping with Hf also improved the irreversibility field.

In the $MgB_2$ system, $J_c$ and $H_{irr}$ are net outcomes of grain connectivity and flux pinning induced by grain boundaries and precipitates. In Ne-ion irradiated $MgB_2$ samples, we noticed that though there was no appreciable decrease in $T_c$, the resistivity increased and $J_c$ decreased due to the reduction in grain connectivity [15] without appreciable change in $T_c^{(R=0)}$. On the contrary, Hf-doped $MgB_2$ shows a substantial increase in $J_c$ with little change in $T_c^{(onset)}$. The change in $T_c^{(R=0)}$ is due to the increase in transition width $\Delta T_c$. In Hf-doped $MgB_2$, the higher $J_c$ may be due to Hf substitution on Mg site (as it is not very likely to go to B-site) and pinning centres formed by precipitation of HfB and $HfB_2$ which are likely to accumulate in the grain boundary. The grain boundary pinning is the dominant picture in $MgB_2$. The pinning force density $F_p$(= $J_c \times B$) at different temperatures for both samples are shown in Fig. 5. At lower temperatures, $F_p$ of the doped sample is higher compared to the undoped sample. It was seen by Dou et al. [16] that SiC precipitation in $MgB_2$ causes an enhancement of $J_c$. In SiC nanoparticle doping into $MgB_2$, it is explained that C-substitution in B sites deteriorates the crystallinity of $MgB_2$ grains resulting in an enhancement of grain boundary pinning [17, 18]. Analyses based on the grain boundary pinning theory revealed that the origin of $J_c$ enhancement comes from the increase of grain boundary density due to the suppression of grain growth and as well as an enhancement of $f_{gB}$ (grain boundary pinning force) due to the existence of non-superconducting phases [19] like HfB and $HfB_2$ in this case. The idea or estimate of these phases as well as the occupancy of Hf at various sites were obtained from TDPAC analysis as is described below.

**TDPAC Analysis:**

The nuclear interaction of the I=5/2 intermediate state in $^{181}$Ta leads to a splitting with three eigenvalues, $E_1$, $E_2$ and $E_3$ : [20]

$$E_1 = -2r\cos\left(\frac{\varphi}{3}\right)$$

$$E_2 = r\cos\left(\frac{\varphi}{3}\right) - \sqrt{3}\, r\sin\left(\frac{\varphi}{3}\right)$$

$$E_3 = r\cos\left(\frac{\varphi}{3}\right) + \sqrt{3}\, r\sin\left(\frac{\varphi}{3}\right)$$

With $\cos\varphi = q/r^3$

$$r = \mathrm{sign}(q)\sqrt{|p|}$$

$$p = -28(1 + \eta^2/3)$$

$$q = -80(1 - \eta^2)$$

with the asymmetry parameter η of the electric field gradient defined as $\eta = (V_{xx} - V_{yy})/V_{zz}$.

The three precession frequencies are:

$$\omega_1 = (E_2 - E_3)\omega_Q$$

$$\omega_2 = (E_1 - E_2)\omega_Q$$

$$\omega_3 = (E_1 - E_3)\omega_Q$$

with the quadrupole frequency :

$$\omega_Q = \frac{eQV_{zz}}{40\hbar}$$

where Q denotes the quadrupole moment of the I = 5/2 state and $V_{zz}$ is the largest component by magnitude of the EFG tensor. The perturbation function is thus:

$$G_2(t) = a_0 + a_1 cos\omega_1 t + a_2 cos\omega_2 t + a_3 cos\omega_3 t$$

The experimental data were fitted with this function modified using finite distributions of $\omega_Q$. In the present case we have used Lorentzian distributions.

Fig.6 shows the TDPAC spectrum for the Hf-doped $MgB_2$ sample annealed at 1173K. The perturbation function is shown on the left and the corresponding cosine transform on the right. The results of the least square fit of the data are presented in Table-1. There are four frequencies corresponding to four inequivalent sites where Hf is present in $MgB_2$ sample. To identify the sites occupied or the species formed by Hf, the TDPAC study was carried out for a sample of $HfB_x$ prepared under identical conditions as that of $MgB_2$. Fig. 7 shows the TDPAC spectra for $HfB_x$ annealed at 1173K and 1373K. The results obtained from these spectra are presented in Table-2. At both temperatures, the spectrum for $HfB_x$ could be fitted with two frequencies, which indicates that Hf exists in two inequivalent sites in $HfB_x$. A closer look at Table-2 indicates that site-II can be assigned to $HfB_2$ for which our experimental quadrupole frequency agrees with the literature value [21]. From the symmetry of $HfB_2$ (P/6mmm) the asymmetry parameter $\eta$ is expected to be zero. The apparent nonzero $\eta$ is most likely related to the finite linewidth (or finite $\delta$) .

The identification of the site-I is not obvious. However, the comparison with XRD results indicates that this low frequency component is most probably the hafnium monoboride, HfB. The site occupancy of this frequency reduces and that of site-II increases with temperature, indicating the conversion of this species to $HfB_2$. The large $\delta$ of this site compared to $HfB_2$ indicates that this species is nonstoichiometric in nature or there could be partial amorphisation if larger islands are formed . HfB is kinetically stable whereas $HfB_2$ is thermodynamically stable. That is why at the low reaction temperature of 1173K, a considerable amount of HfB is formed along with $HfB_2$, but at the higher temperature of 1373K, the amount of $HfB_2$ increases as it is evident from Table-2.

Referring back to Table-1, we find that the same HfB species is present and this is a major component among the Hf species in the Hf-doped $MgB_2$ sample annealed at 1173K. On the other hand, $HfB_2$ is a minority phase at this temperature. There are two more sites where Hf resides which correspond to the remaining two frequencies ($\omega_Q$=59.5 Mrad/s and 248.9 Mrad/s). It is highly probable that Hf can go to both Mg and B sites, which gives rise to these two frequencies. The justification for this assumption stems from $V_{zz}$ of these two lattice sites. Literature data on the EFG [22,23] from an NMR study at these two sites using $^{25}$Mg and $^{12}$B, respectively, showed that the EFG at the B site is nearly five times larger than that at the Mg site. Our results presented in Table-2 also have a similar ratio (>4.0). However, the absolute values of the EFG for $^{181}$Ta probes compared to other probes [24] are high and hence the same trend is expected for Mg or B sites. A logical conclusion would be that Hf occupies both Mg and B positions. The site occupancies of the two sites are not simply statistical but a consequence of the chemical potential between Ta and the surrounding atoms. However, considering the ionic radii of the species involved (Mg(2+), Hf(4+), and B(-1) being 0.074, 0.071 and 0.09nm, respectively), it is highly probable that Hf predominantly occupies the Mg site. Moreover, from the results of $T_c$ (in particular $T_c^{(onset)}$) it is obvious that Hf occupies B site to a less extent than Mg site. Thus the TDPAC results mentioned in Table-1 clearly indicate that the site-II corresponds to Mg and site-III to B. Considering the site occupancy it is quite obvious that the largest component HfB may precipitate and act as the most effective pinning centre for the enhancement of $J_c$. The other two substitutions, i.e. Hf in Mg and B sites play minor roles in the enhancement of $H_{c2}$.

**Conclusion:**

We have observed a considerable enhancement of $J_c$, $H_{irr}$ and also $H_{c2}$ in 1% Hf-doped $MgB_2$, contrary to charged particle irradiation. There was little change in $T_c^{(onset)}$ rendering Hf-doping an effective means to enhance $J_c$ without losing superconducting properties. The enhancement of $J_c$ is caused primarily by HfB and $HfB_2$ precipitations at grain boundary, whereas the increase of $H_{c2}$ points towards substitution of Hf at B and Mg sites Substitution at B site would affect 2D σ-bands and hence $T_c^{(onset)}$, whereas that at Mg site would affect the 3D π-bands and hence increase the resistivity. The insignificant change of $T_c^{(onset)}$ points towards Hf occupying Mg site dominantly. There has been a considerable increase in resistivity as well as the transition width, implying a reduction of grain connectivity, too. The TDPAC spectra of Hf-doped $MgB_2$ revealed four inequivalent sites which were compared with the reference Hf-boride annealed at the temperature of formation of $MgB_2$ and at a higher temperature. From XRD it is evident that at the temperature of formation of $MgB_2$, the major species formed is HfB, which was converted to the thermodynamically stable phase $HfB_2$ at higher temperatures. Thus, HfB precipitates formed in Hf-doped $MgB_2$ act as effective pinning centres for the enhancement of $J_c$, the grain boundary pinning mechanism being dominant. There is an indication of partial occupation of both Mg and B sites by Hf as evident from the relative EFG of other two components. At 1% doping, this partial occupation of B site if at all, does not destroy the superconductivity. Higher levels of doping may be attempted to investigate the occupancy of Hf at B site and its effects. But the 1% level of Hf-doping is beneficial from the point of view of superconductivity. We are planning to carry out the synthesis of Hf-doped $MgB_2$ at temperatures higher than 1173 K and observe $J_c$ to further elucidate the role of HfB/$HfB_2$ for the enhancement of $J_c$.

**Acknowledgement**:

Authors sincerely thank Dr. S.V. Thakare, Radiopharmaceuticals Division, BARC, Dr. Ms. P. Mukherjee, Material Group, VECC and Dr. K. Krishnan, Fuel Chemistry Division, BARC for their help in the experiment. They are also thankful to Dr. V.K. Manchanda, Head Radiochemistry Division, BARC and Dr. S.K. Aggarwal, Head, Fuel Chemistry Division, BARC for their keen interest in this work. One of the authors (AT) thanks CSIR for the research grant.

Table1. TDPAC parameters: quadrupole frequency $\omega_Q$, asymmetry parameter $\eta$, Lorentzian frequency distribution $\delta$, and site occupancy for $^{181}Hf$ in $MgB_2$

Annealing temp: 1173K

| **Site** | **$\omega_Q$ (Mrad/s)** | **$\eta$** | **$\delta$ (%)** | **Site %** |
|---|---|---|---|---|
| I | 32.3(5) | 0.41(2) | 8.2 (5) | 49 |
| II | 59.5(9) | 0.17(6) | 12.1(1.7) | 26 |
| III | 248.9(7) | 0.17(5) | 8.0(1) | 16 |
| IV | 112.7( 5) | 0.10(1 ) | 1.9(7) | 9 |

Table 2. TDPAC parameters: quadrupole frequency $\omega_Q$, asymmetry parameter $\eta$, Lorentzian frequency distribution $\delta$, and site occupancy for $^{181}Hf$ in $HfB_x$

| **Site** | **$\omega_Q$ (Mrad/s)** | **$\eta$** | **$\delta$ (%)** | **Site %** |
|---|---|---|---|---|
| **Annealing Temperature 1173K** | | | | |
| I | 34.1(2) | 0.45(4) | 8.2(3) | 34 |
| II | 113.2(4) | 0.1(1) | 2.4(4) | 66 |
| **Annealing Temperature 1373K** | | | | |
| I | 34.0(2) | 0.41(6) | 2.6(4) | 10 |
| II | 113.3(1) | 0.10(1) | 0.3(1) | 90 |

**Figure Captions:**

Fig. 1 XRD patterns of (a) Hf-doped $MgB_2$, (b) $HfB_2$ annealed at 1173K

Fig. 2. Resistivity versus temperature for pristine $MgB_2$ and Hf-doped $MgB_2$. Inset shows superconducting transitions for the samples to indicate the transition width.

Fig. 3. Upper critical field ($H_{c2}$) of undoped and Hf-doped $MgB_2$. Inset shows irreversibility field line $H_{irr}(T)$ of the samples. The behaviour of $H_{c2}$ and $H_{irr}$ is similar.

Fig. 4. Critical current density of undoped and Hf-doped $MgB_2$ at (a) 2 K and (b) 20 K.

Fig. 5. Pinning force density at various temperatures of undoped and Hf-doped $MgB_2$.

Fig. 6. TDPAC spectrum (left) and cosine-transform (right) of Hf-doped $MgB_2$ annealed at 1173 K. The points correspond to the experimental data and line to the fit of the experimental data.

Fig. 7. Top: TDPAC spectrum (left) and cosine-transform (right) of $HfB_x$ annealed at 1173 K.

Bottom: annealed at 1373 K. The points correspond to the experimental data and line to the fit of the experimental data.

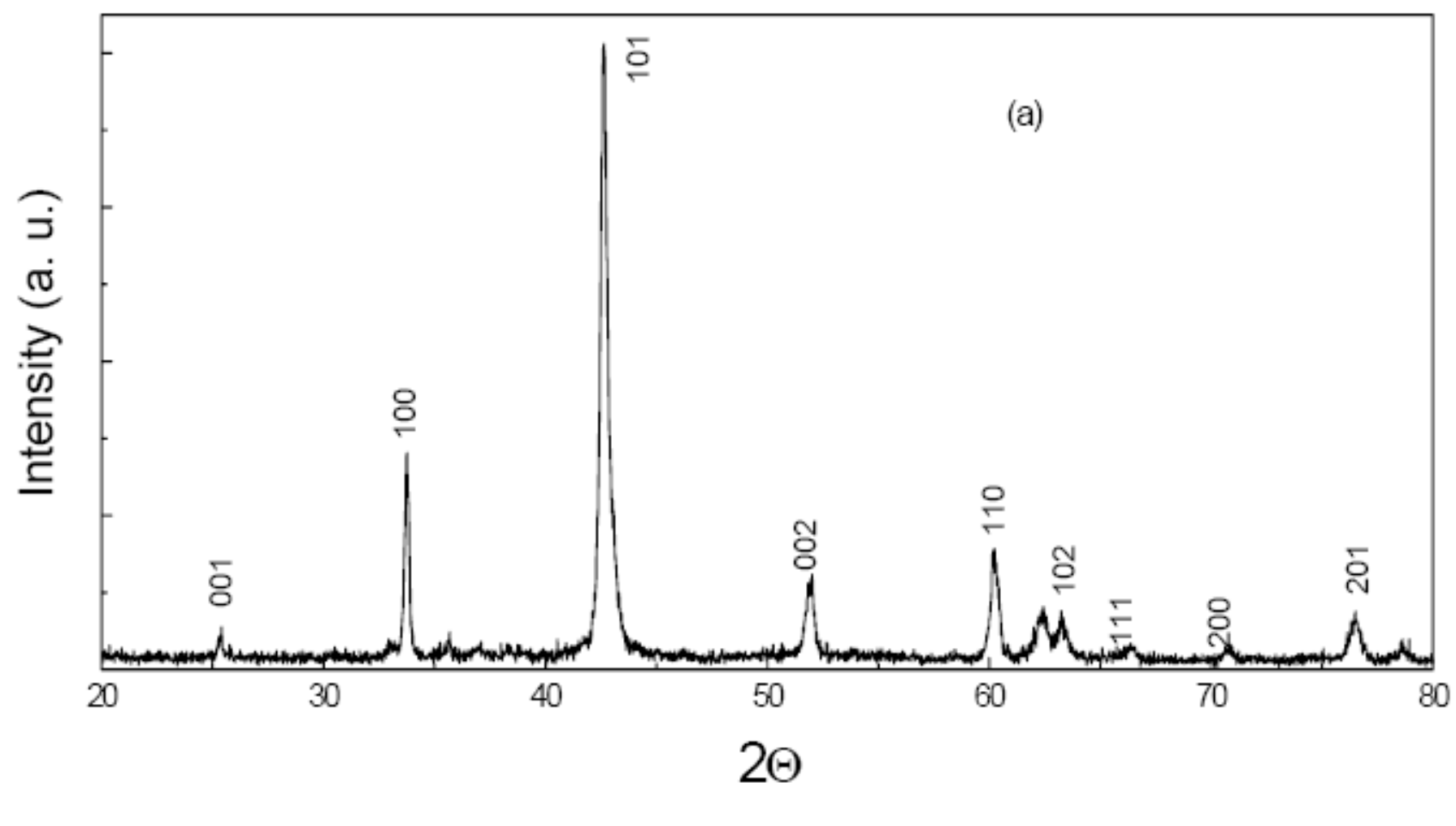
Intensity (a. u.)
2Θ
(a)
001
100
101
002
110
102
111
200
201
20
30
40
50
60
70
80

Fig 1 (a)

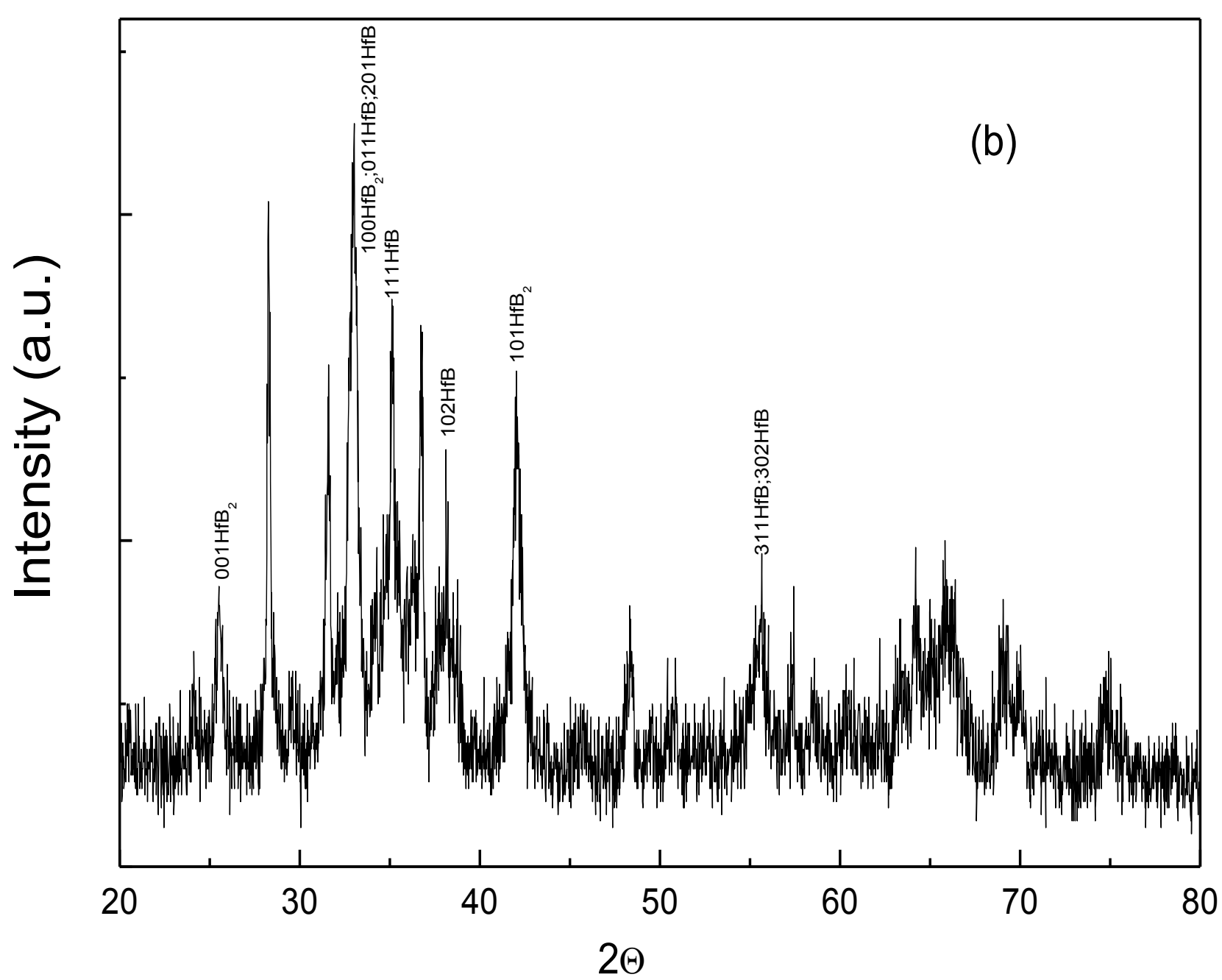
Intensity (a.u.)
2Θ
(b)
001HfB2
100HfB2;011HfB;201HfB
111HfB
102HfB
101HfB2
311HfB;302HfB
20
30
40
50
60
70
80

Fig 1 (b)

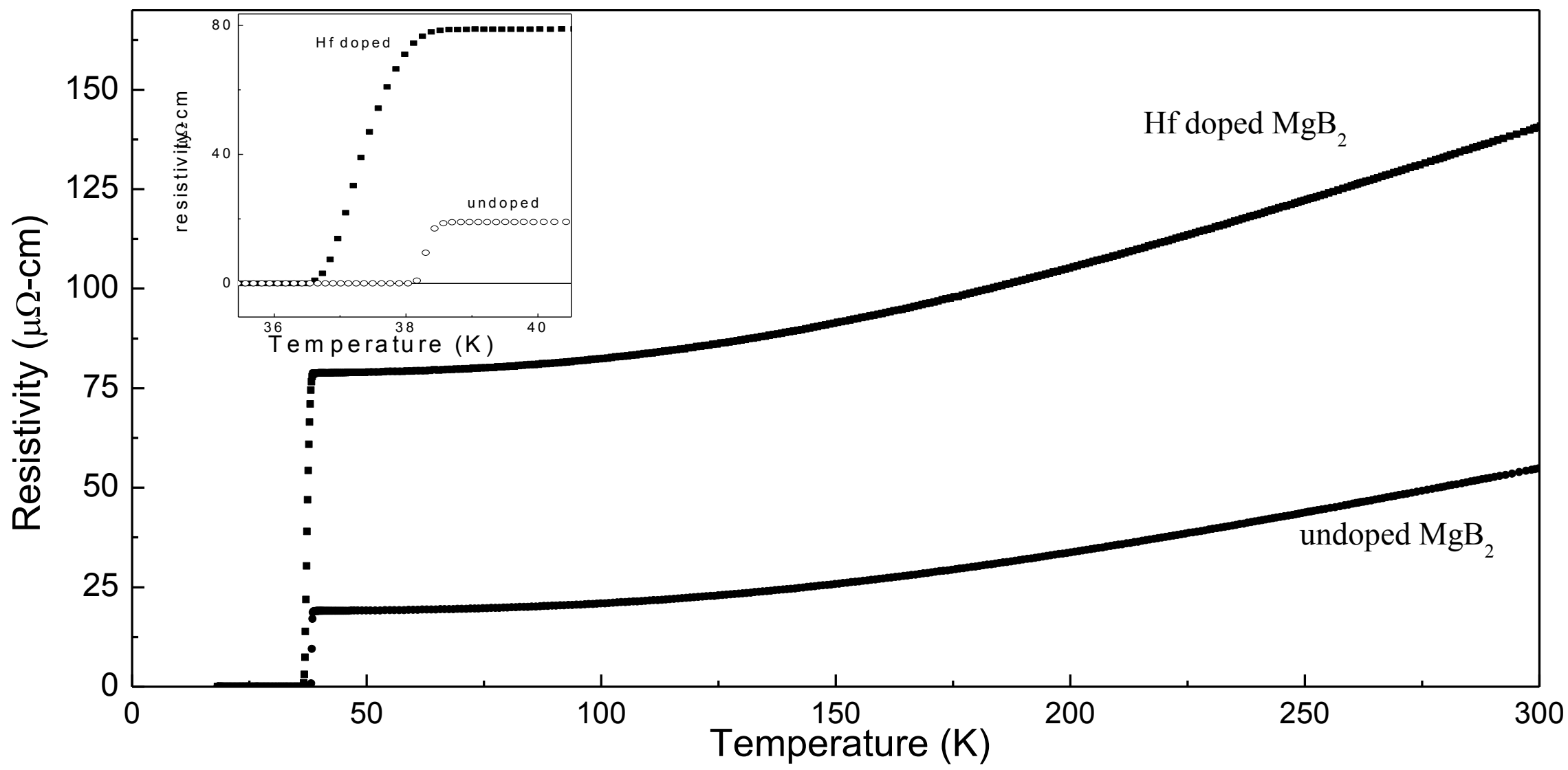
Resistivity (μΩ-cm)
Temperature (K)
Hf doped MgB2
undoped MgB2
Hf doped
undoped
resistivity μΩcm
Temperature (K)

Fig. 2

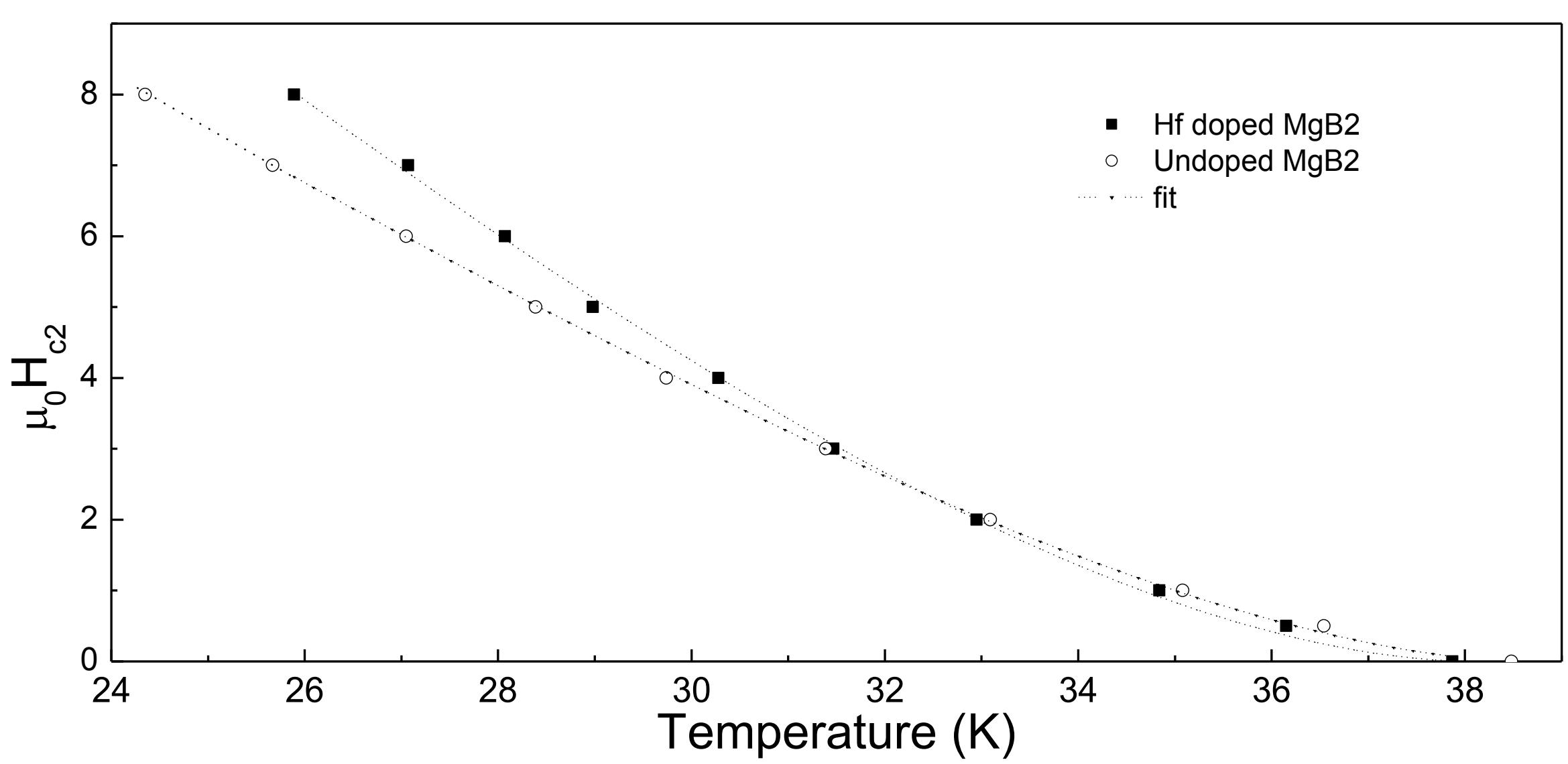
$\mu_0 H_{c2}$
Temperature (K)
Hf doped MgB2
Undoped MgB2
fit

Fig. 3

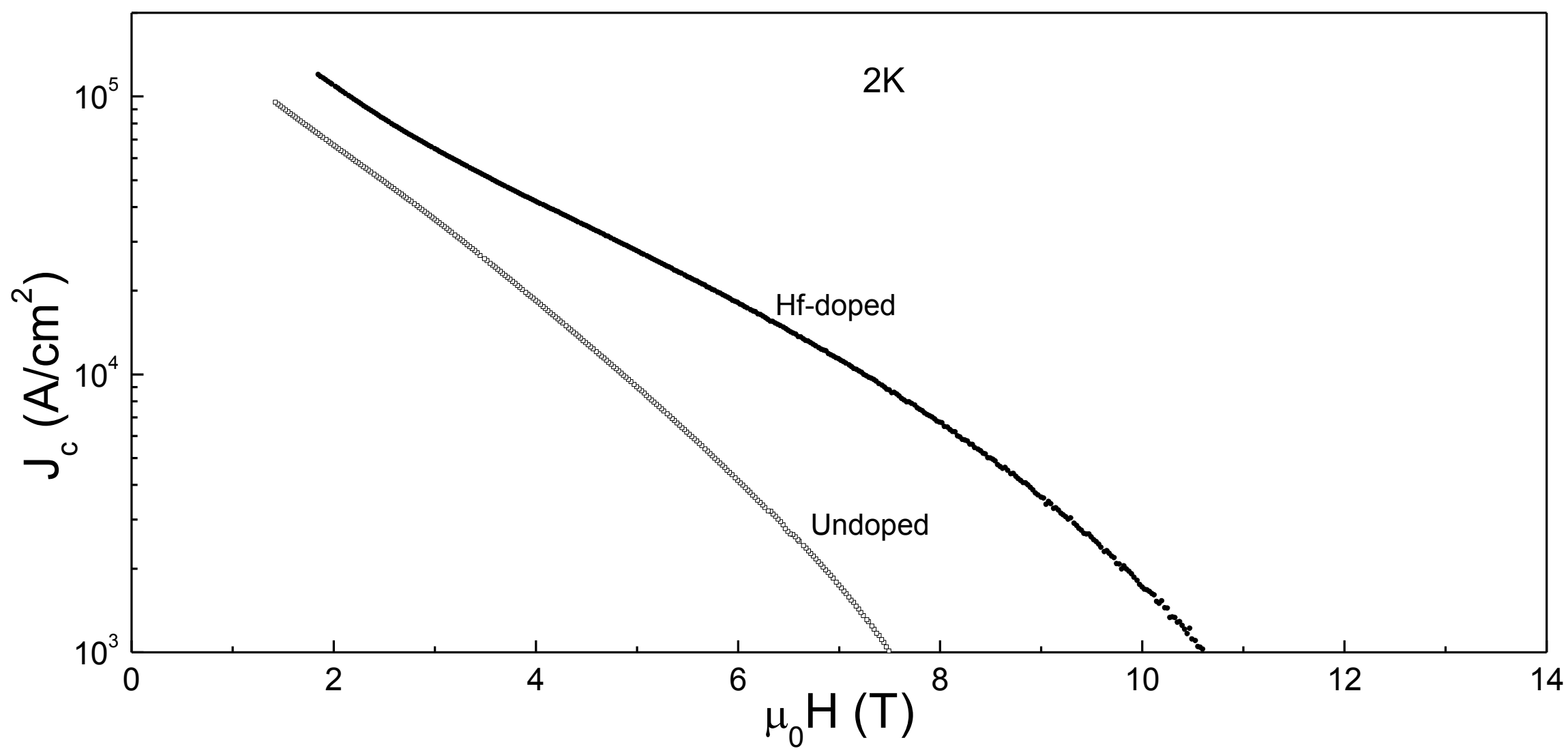
2K
Hf-doped
Undoped
J_c (A/cm^2)
μ_0H (T)

Fig. 4a

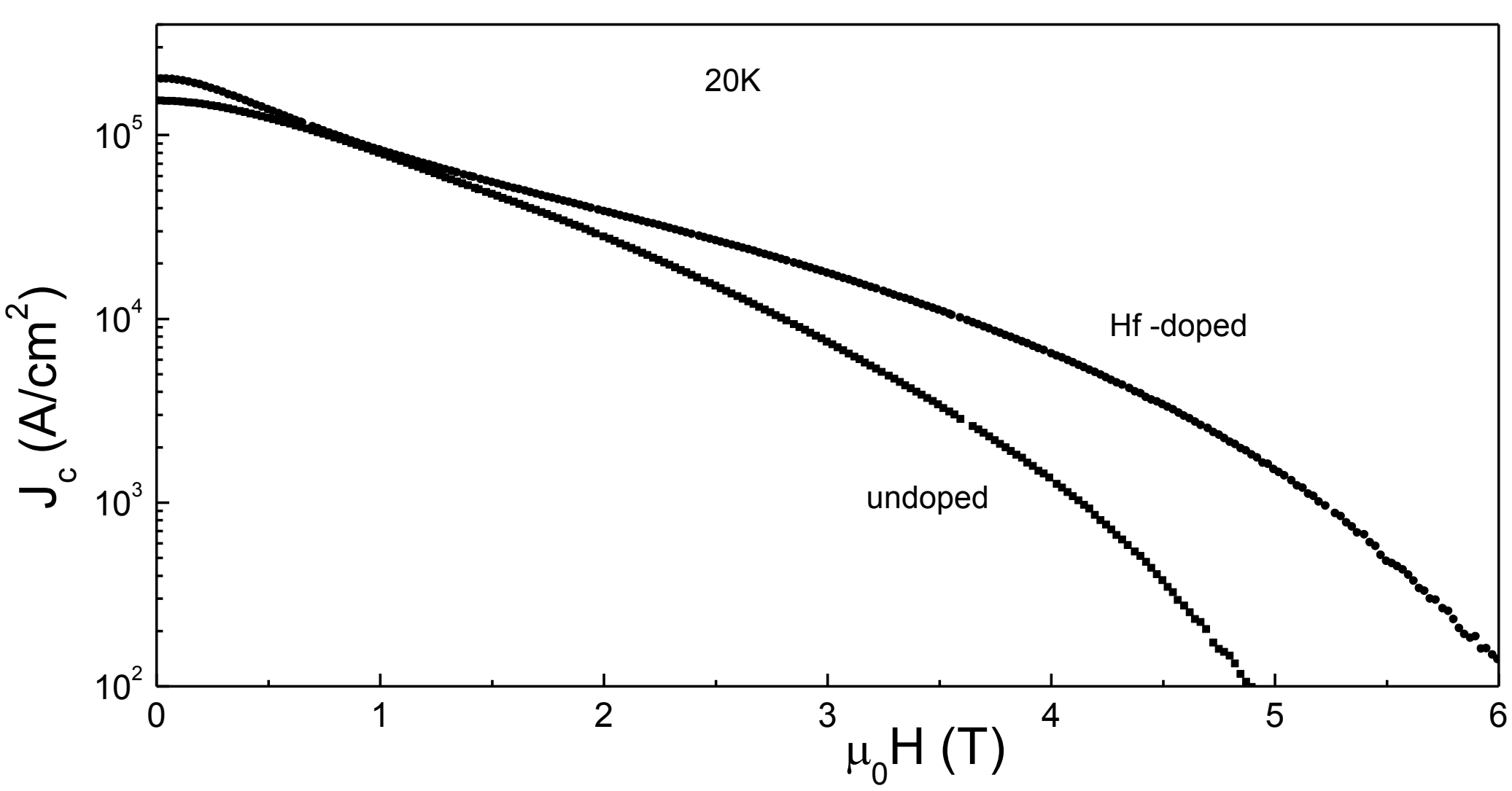
20K
Hf -doped
undoped
J_c (A/cm^2)
μ_0H (T)

Fig 4 (b)

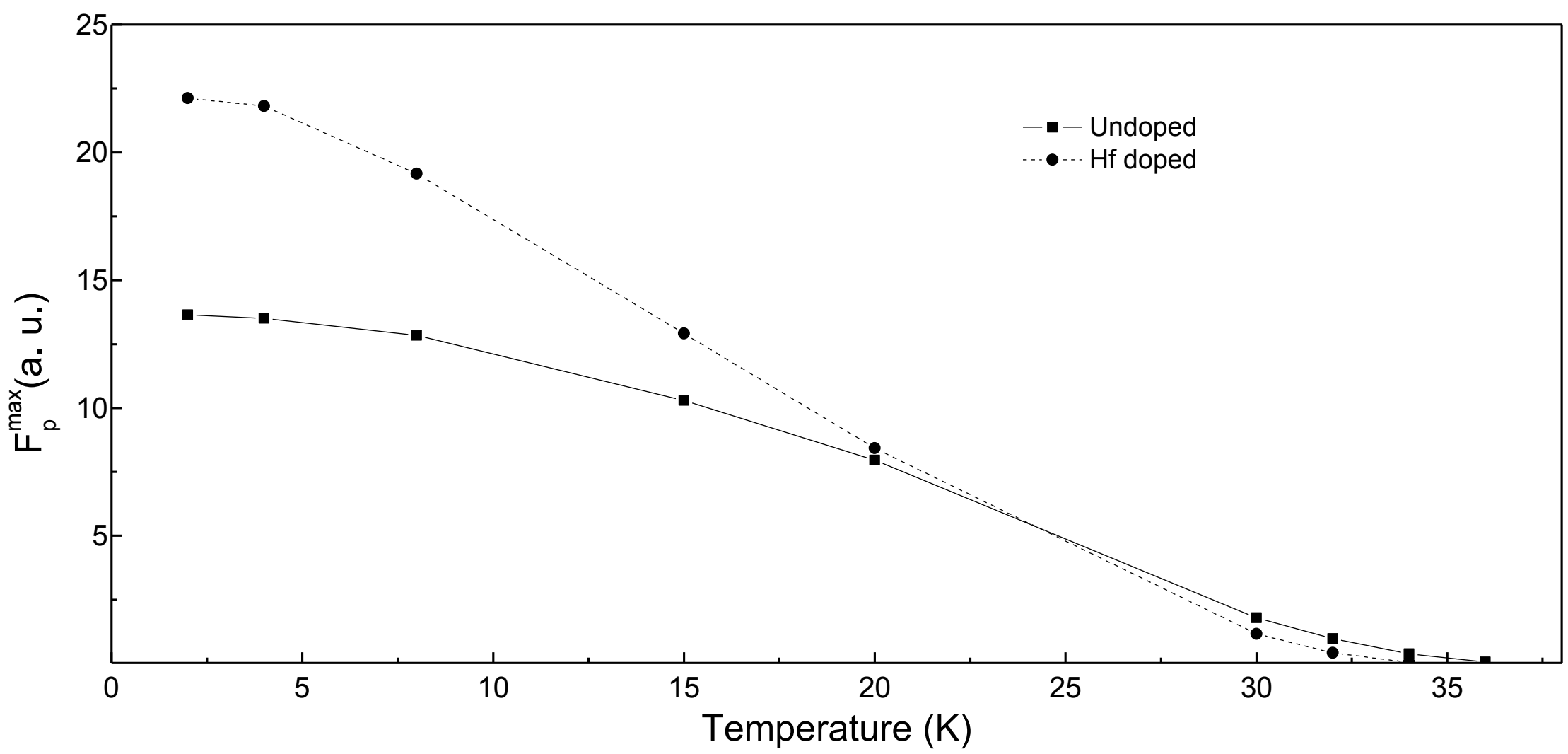

Fig. 5

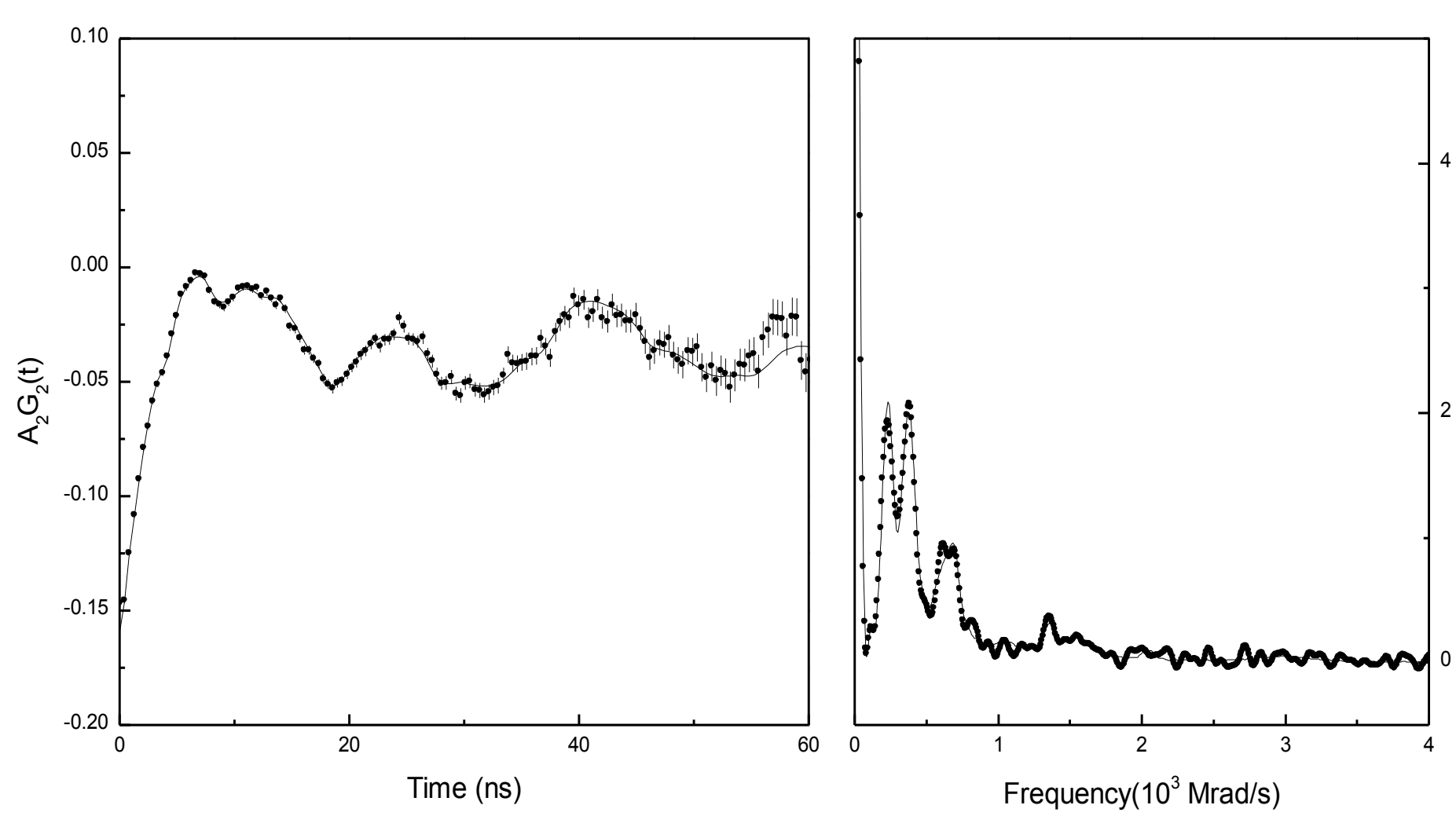

Fig 6.

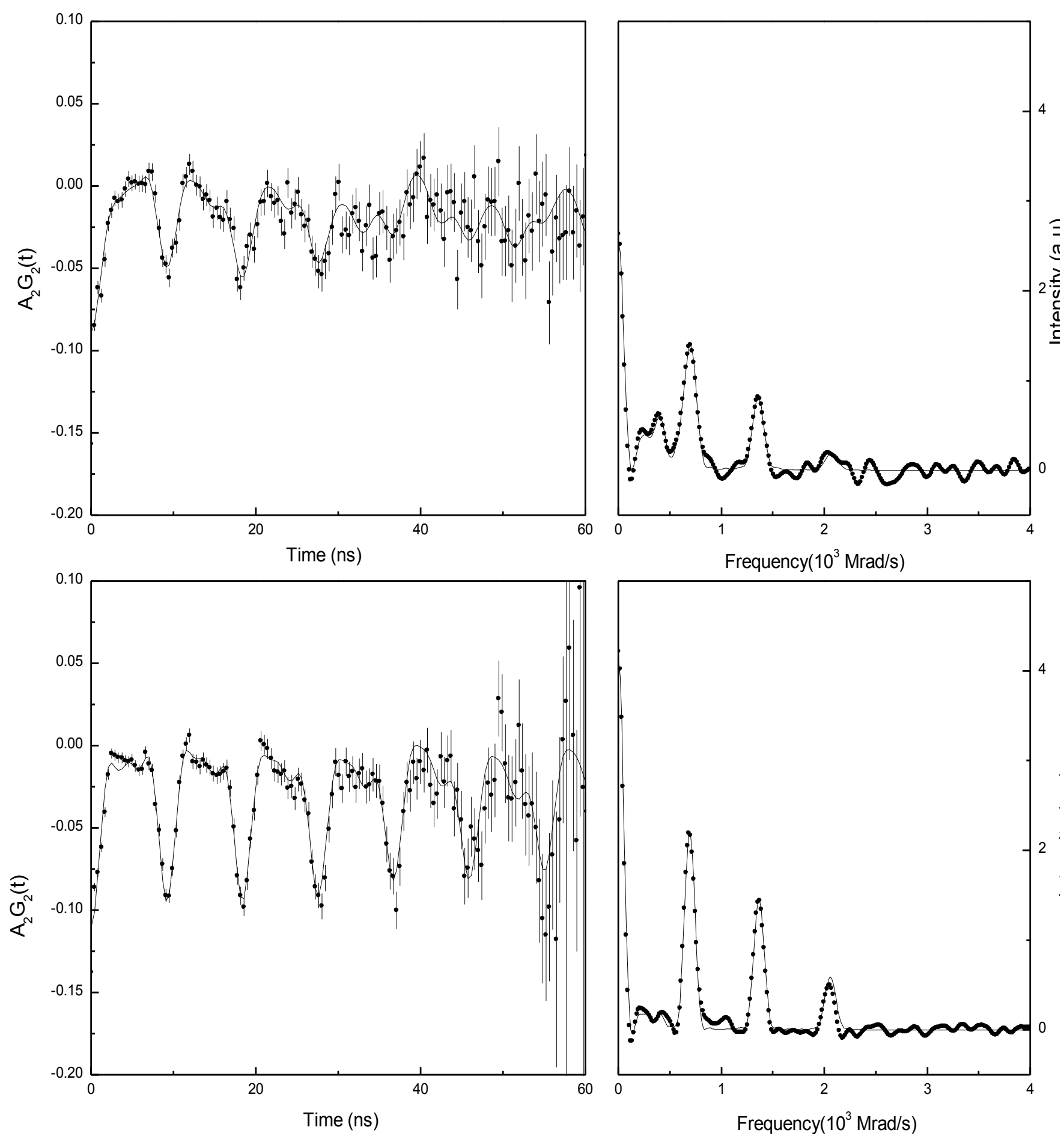

Fig 7.